\documentclass{elsart}

\usepackage{natbib}

 \usepackage{graphics}

\def\mb#1{\setbox0=\hbox{$#1$}\kern-.025em\copy0\kern-\wd0
\kern-0.05em\copy0\kern-\wd0\kern-.025em\raise.0233em\box0}

\begin{document}

\begin{frontmatter}



\title{Statistical mechanics of the shallow-water system with
a prior potential vorticity distribution}


\author{P.H. Chavanis and B. Dubrulle}

\address{$^1$ Laboratoire de Physique Th\'eorique (UMR 5152), Universit\'e Paul
Sabatier, Toulouse, France\\
$^2$ DRECAM/SPEC/CEA Saclay, and CNRS (URA2464), F-91190 Gif sur
Yvette Cedex, France\\}
\ead{chavanis@irsamc.ups-tlse.fr; Berengere.Dubrulle@cea.fr}

\begin{abstract}
We adapt the statistical mechanics of the shallow-water equations to
the case where the flow is forced at small scales. We assume that
the statistics of forcing is encoded in a prior potential vorticity
distribution which replaces the specification of the Casimir
constraints in the case of freely evolving flows. This determines a
generalized entropy functional which is maximized by the
coarse-grained PV field at statistical equilibrium. Relaxation
equations towards equilibrium are derived which conserve the robust
constraints (energy, mass and circulation) and increase the
generalized entropy.
\end{abstract}


\end{frontmatter}

\section{Introduction}

Two-dimensional turbulent flows with high Reynolds numbers have the
striking property of organizing spontaneously into large-scale
coherent structures such as jets and vortices. Jovian atmosphere
shows a wide diversity of structures: Jupiter's great red spot,
white ovals, brown barges,... A good hydrodynamical model to
describe jovian atmosphere is provided by the Shallow-Water (SW)
equations. A statistical theory of the SW system has been developed
recently by \citet{sw}. This extends the statistical mechanics of
the incompressible 2D Euler equation proposed by \citet{miller} and
\citet{rs} to the case of compressible flows. It is however assumed
that the flow is weakly compressible (small Mach number) so that the
effect of waves is not dominant. Therefore, large-scale coherent
vortices can form and persist for a long time. After a phase of
chaotic mixing (violent relaxation), the system is expected to reach
an equilibrium state which corresponds to the most mixed state
consistent with the constraints imposed by the dynamics.
Mathematically, it is obtained by maximizing a mixing entropy while
conserving energy, mass, circulation and all the higher moments of
the PV vorticity (Casimirs). One difficulty with this approach is
that the moments of PV depend on the resolution at which the PV
field is considered so that the prediction can change accordingly.
Furthermore, geophysical flows are usually forced and dissipated at
small-scales so that the conservation of all the Casimirs is
abusive. In an attempt to solve these problems, \citet{ellis} have
proposed to fix a prior vorticity distribution instead of the
Casimirs. It is assumed that this global distribution of vorticity
is generated by the small-scale forcing so it must be given as an
{\it input} in the statistical theory. In this approach, only the
conservation of the robust constraints $(E,\Gamma)$ is taken into
account and the effect of the small-scale forcing is encapsulated in
a prior vorticity distribution or in a generalized entropy. This
approach has been further developed in \citet{physicad} and
\citet{all} in the context of the quasi-geostrophic (QG) equations.
In this paper, we show how it can be extended to the case of the
Shallow-Water (SW) equations. Since this is only a slight, but
interesting, variant with respect to the un-forced case, we shall
mostly refer to the study of \citet{sw} for technical details,
without repeating all the steps of the derivation.

\section{The shallow-water equations}
\label{sec_sa}

The dynamical evolution of a thin fluid layer with thickness
$h(x,y,t)$ and velocity field ${\bf u}=(u,v)(x,y,t)$ submitted to a
gravity acceleration $g$ on a rotating planet is governed by the
shallow water equations
\begin{equation}
{\partial h\over\partial t}+\nabla \cdot (h {\bf u})=0,
\label{sa1}
\end{equation}
\begin{equation}
{\partial {\bf u}\over\partial t}+({\bf u}\cdot\nabla){\bf u}=-g\nabla h-2{\bf\Omega}\times{\bf u}.
\label{sa2}
\end{equation}
The first equation can be viewed as an equation of continuity and the
second equation as the Euler equation (in a rotating frame
${\bf\Omega}$) for a barotropic fluid with pressure $p={1\over 2}g
h^{2}$. The Euler equation can be rewritten
\begin{equation}
{\partial {\bf u}\over\partial t}+({\mb \omega}+2{\bf\Omega})\times {\bf u}=-\nabla B,
\label{sa3}
\end{equation}
where ${\mb \omega}=\omega{\bf e}_{z}=\nabla\times{\bf u}$ is the vorticity and where we have introduced the Bernouilli function
\begin{equation}
B={{\bf u}^{2}\over 2}+gh.
\label{sa4}
\end{equation}
The potential vorticity (PV)
\begin{equation}
q={\omega+2\Omega\over h},
\label{sa5}
\end{equation}
is conserved for each fluid parcel, i.e.
\begin{equation}
{dq\over dt}={\partial q\over\partial t}+{\bf u}\cdot \nabla q=0.
\label{sa6}
\end{equation}
Each mass element $h d{\bf r}$ is also conserved in the course of
the evolution. This implies that the shallow-water equations conserve
the PV moments
\begin{equation}
\Gamma_{n}=\int q^{n}h d{\bf r}.
\label{sa7}
\end{equation}
The moments $n=0,1,2$ are, respectively, the total mass $M$, the
circulation $\Gamma$ and the PV enstrophy $\Gamma_{2}$. The
energy
\begin{equation}
E=\int h {{\bf u}^{2}\over 2}d{\bf r}+{1\over 2}\int g h^{2}d{\bf r},
\label{sa8}
\end{equation}
involving a kinetic and a potential part is also conserved. It is convenient to use a Helmholtz decomposition of the momentum $h{\bf u}$ into a purely rotational and purely divergent part
\begin{equation}
h{\bf u}=-{\bf e}_{z}\times\nabla\psi+\nabla\phi.
\label{sa9}
\end{equation}

For any stationary solution, the mass conservation (\ref{sa1}) reduces
to $\nabla\cdot (h{\bf u})=0$ so that
\begin{equation}
h{\bf u}=-{\bf e}_{z}\times \nabla\psi,
\label{sa10}
\end{equation}
where $\psi$ is the stream-function. Then Eq. (\ref{sa6}) reduces to ${\bf u}\cdot \nabla q=0$ which implies that $q=f(\psi)$. Finally, Eq. (\ref{sa3}) reduces to
\begin{equation}
({\mb \omega}+2{\bf\Omega})\times {\bf u}=-\nabla B.
\label{sa11}
\end{equation} Taking the scalar product with ${\bf u}$, we obtain ${\bf u}\cdot \nabla B=0$ so that  $B=B(\psi)$. Then, substituting Eq. (\ref{sa10}) in Eq. (\ref{sa11}) we obtain
\begin{equation}
q=-{dB\over d\psi}.
\label{sa12}
\end{equation}
The SW equations admit an infinite class of stationary solutions, specified by the relations $B=B(\psi)$ and  $q=f(\psi)=-B'(\psi)$. They are determined by the two coupled partial differential differential equations for $\psi$ and $h$
\begin{equation}
-{\Delta\psi\over h^{2}}+{2\Omega\over h}+{1\over h^{3}}\nabla\psi\cdot\nabla h=-{dB\over d\psi},
\label{sa13}
\end{equation}
\begin{equation}
{(\nabla\psi)^{2}\over 2 h^{2}}+gh=B(\psi).
\label{sa14}
\end{equation}
This formulation of the SW equations in terms of $(h,\psi)$ variables
has been introduced in  \citet{sw}.

\section{The equilibrium statistical mechanics}

\subsection{Freely evolving flows}
\label{sec_free}

The SW equations are known to develop a complicated mixing process
which ultimately leads to the emergence of a large-scale coherent
structure, typically a jet or a vortex. One question of fundamental
interest is to understand and predict the structure and the stability
of these equilibrium states. This can be achieved by using statistical
mechanics arguments. The idea is to replace the deterministic
description of the flow $q({\bf r},t)$ by a probabilistic description
where $\rho({\bf r},\sigma,t)$ gives the density probability of
finding the PV level $q=\sigma$ in ${\bf r}$ at time $t$ (it satisfies
the local normalization condition $\int \rho({\bf
r},\sigma)d\sigma=1$). The observed (coarse-grained) PV field is then
expressed as $\overline{q}({\bf r},t)=\int \rho\sigma d\sigma$. To
apply the statistical theory, one must first specify the constraints
attached to the SW equations. The mass $M=\int h d{\bf r}$,
circulation $\Gamma=\int
\overline{q}h d{\bf r}$ and energy  $E=\int h {{\bf u}^{2}\over 2}d{\bf r}+{1\over 2}\int g h^{2}d{\bf r}$ will be called {\it robust
constraints} because they can be expressed in terms of the
coarse-grained field. These integrals can be calculated at any time
from the coarse-grained field and they are conserved by the
(macroscopic) dynamics. By contrast, the Casimir invariants $I_{f}=\int
\overline{f(q)}hd{\bf r}$, or equivalently the fine-grained
moments of the vorticity $\Gamma_{n>1}^{f.g.}=\int
\overline{q^{n}}h d{\bf r}=\int \rho\sigma^{n}d\sigma h d{\bf r}$,
will be called {\it fragile constraints} because they must be
expressed in terms of the fine-grained PV. Indeed, the moments of the
coarse-grained PV $\Gamma_{n>1}^{c.g}=\int
\overline{q}^{n}hd{\bf r}$ are {\it not} conserved since
$\overline{q^{n}}\neq \overline{q}^{n}$ (part of the coarse-grained
moments goes into fine-grained fluctuations). Therefore, the moments
$\Gamma_{n>1}^{f.g.}$ must be calculated from the fine-grained field
$q({\bf r},t)$ or from the initial conditions, i.e. before the PV
vorticity has mixed. Since we often do not know the initial conditions
nor the fine-grained field, the Casimir invariants often appear as
``hidden constraints'' \citep{super}.

The statistical theory of Miller and Robert-Sommeria-Chavanis for the
2D Euler equations and the SW equations is based on several
assumptions: (i) it is assumed that the flow is freely evolving
without small-scale forcing and dissipation. (ii) it is assumed that
we know the initial conditions (or equivalently the value of all the
Casimirs) in detail. (iii) it is assumed that mixing is efficient and
that the evolution is ergodic so that the system will reach at
equilibrium the most probable (most mixed) state. Within these
assumptions\footnote{Some attempts have been proposed to go beyond the
assumptions (ii) and (iii) of the statistical theory. For example,
\citet{jfm1} consider a {\it strong mixing limit} in which only the
first moments of the vorticity are relevant instead of the whole set
of Casimirs. On the other hand, \citet{jfm2} introduce the concept of
{\it maximum entropy bubbles} (or restricted equilibrium states) in
order to account for situations where the evolution of the flow is not
ergodic in the whole available domain but only in subdomains. A 2D
turbulent flow is therefore viewed as an ensemble of isolated vortices
which can be seen as ``maximum entropy bubbles'' separated by un-mixed
flow. In 2D decaying turbulence, these isolated vortices result from
previous mergings and they are expected to correspond to statistical
equilibrium states
\citep{laval}. In fact, because of {\it incomplete relaxation}, they
may just well be particular stable stationary solutions of the 2D
Euler equation that are incompletely mixed.}, the statistical
equilibrium state of the SW system is obtained by maximizing the
mixing entropy
\begin{equation}
S[\rho]=-\int \rho\ln\rho h d{\bf r}d\sigma,
\label{free1}
\end{equation}
while conserving the energy, the circulation, the mass  and {\it all the Casimirs}. We write the variational principle in the form
\begin{equation}
\delta S-\beta\delta E-\alpha\delta\Gamma-\gamma\delta M-\sum_{n>1}\alpha_{n}\delta\Gamma_{n}^{f.g.}-\int\zeta({\bf r})\delta\biggl (\int \rho d\sigma\biggr )hd{\bf r}=0.
\label{free2}
\end{equation}
In the present point of view, all the constraints are treated on the
same footing. In particular, the moments $\Gamma_{n}^{f.g.}$ are
treated microcanonically and we must ultimately relate the Lagrange
multipliers $\alpha_{n}$ to the constraints $\Gamma_{n}^{f.g.}$.

\subsection{Prior distribution and relative mixing entropy}
\label{prior}

In the approach of \citet{sw}, it is assumed that the system is
strictly described by the SW equations so that the conservation of all
the Casimirs has to be taken into account.  However, in geophysical
situations, the flows are forced and dissipated at small scales (due
to convection in the jovian atmosphere) so that the conservation of
the Casimirs is destroyed.  \citet{ellis} have proposed to treat these
situations by fixing the conjugate variables $\alpha_{n>1}$ instead of
the fragile moments $\Gamma_{n>1}^{f.g.}$. If we view the PV levels as
species of particles, this is similar to fixing the chemical
potentials instead of the total number of particles in each
species. Therefore, the idea is to treat the fragile constraints {\it
canonically}, whereas the robust constraints $(E,\Gamma,M)$ are still
treated {\it microcanonically}. This point of view has been further
developed in \citet{physicad} and \citet{all} in the QG model and this
approach is extended here to the SW system. The relevant
thermodynamical potential is obtained from the mixing entropy
(\ref{free1}) by using a Legendre transform with respect to the fragile
constraints:
\begin{equation}
S_{\chi}=S-\sum_{n>1}\alpha_{n}\Gamma_{n}^{f.g.}.
\label{p1}
\end{equation}
Explicating the fine-grained moments, we obtain
\begin{equation}
S_{\chi}=-\int \rho \biggl\lbrack \ln\rho+\sum_{n>1}\alpha_{n}\sigma^{n}\biggr\rbrack h d{\bf r}d\sigma.
\label{p2}
\end{equation}
Introducing the function
\begin{equation}
\label{p3}
\chi(\sigma)\equiv{\rm exp}\biggl\lbrace -\sum_{n>1}\alpha_{n}\sigma^{n}\biggr\rbrace.
\end{equation}
we get
\begin{equation}
S_{\chi}[\rho]=-\int \rho \ln\biggl\lbrack {\rho\over\chi(\sigma)}\biggr\rbrack h d{\bf r}d\sigma,
\label{p4}
\end{equation}
which has the form of a relative entropy. The function $\chi(\sigma)$
is interpreted as a prior vorticity distribution. It is a global
distribution of PV fixed by the small-scale forcing. We shall assume
that this function is {\it imposed} by the small-scale forcing so it
must be regarded as {\it given}.

Assuming ergodicity, the statistical equilibrium state is now obtained
by maximizing the relative entropy $S_{\chi}$ at fixed energy $E$,
circulation $\Gamma$ and mass $M$ (robust constraints). The
conservation of the Casimirs has been replaced by the specification of
the prior $\chi(\sigma)$. From that point, we can repeat the
calculations of \citet{sw} with almost no modification. The only
difference is that we regard the $\alpha_{n}$ as given. Writing
$\delta S_{\chi}-\beta\delta E-\alpha\delta\Gamma-\gamma\delta M=0$,
and accounting for the normalization condition $\int\rho d\sigma=1$,
we find (i) that the statistical equilibrium state is a stationary
solution of the SW equation (ii) that the detailed distribution of PV
levels is given by the Gibbs state
\begin{equation}
\rho({\bf r},\sigma)={1\over Z}\chi(\sigma)e^{-(\beta\psi+\alpha)\sigma},
\label{p5}
\end{equation}
where the partition function
\begin{equation}
Z=\int\chi(\sigma)e^{-(\beta\psi+\alpha)\sigma}d\sigma,
\label{p6}
\end{equation}
is determined by the local normalization condition. The distribution
of the fluctuations of PV vorticity is the product of a universal
Boltzmann factor by a non-universal function $\chi(\sigma)$ fixed by
the forcing. This is the same formula as in \citet{sw} except that in
the present formalism $\chi(\sigma)$ must be regarded as given {\it a
priori} while in \citet{sw} it was a function of the Lagrange
multipliers $\alpha_{n}$ that had to be related {\it a posteriori} to
the fine-grained moments imposed by the initial conditions.  The
equilibrium coarse-grained PV is
\begin{equation}
\overline{q}={\int \chi(\sigma)\sigma e^{-(\beta\psi+\alpha)\sigma} d\sigma\over \int \chi(\sigma) e^{-(\beta\psi+\alpha)\sigma} d\sigma}=-{1\over\beta}{d\ln Z\over d\psi}=F(\beta\psi+\alpha)=f(\psi).
\label{p7}
\end{equation}
The coarse-grained vorticity (\ref{p7}) can be viewed as a sort of
{\it superstatistics} \citep{super} as it is expressed as a superposition of
Boltzmann factors (on the fine-grained scale) weighted by a
non-universal function $\chi(\sigma)$. We note that the $\overline{q}-\psi$ relationship
predicted by the statistical theory can take a wide diversity of forms
(usually non-Boltzmannian) depending on the prior $\chi(\sigma)$. The
function $F$ is entirely specified by the prior PV distribution according to
\begin{equation}
\label{p8}
 F(\Phi)=-(\ln\hat\chi)'(\Phi),\qquad {\rm with}\quad \hat{\chi}(\Phi)=\int_{-\infty}^{+\infty}\chi(\sigma)
e^{-\sigma\Phi}d\sigma.
\end{equation}
Differentiating Eq. (\ref{p7}) with respect to $\psi$, we find that
\begin{equation}
\label{p9}
\overline{q}'(\psi)=-\beta q_{2}, \qquad q_{2}=\overline{q^{2}}-\overline{q}^{2}\ge 0.
\end{equation}
This relation relates the slope of the $\overline{q}=f(\psi)$ relationship to the local centered variance $q_{2}(\psi)$ of the PV distribution.  This can also be written $F'(\Phi)=-q_{2}(\Phi)\le 0$ so that $F$ is a decreasing
function.  Therefore, the statistical theory predicts that
the coarse-grained  field is a {\it
stationary solution} of the SW equation and that the
$\overline{q}-\psi$ relationship is a {\it monotonic} function
which is increasing at {negative temperatures} $\beta<0$ and
decreasing at positive temperatures $\beta>0$. We note that, according to Eqs. (\ref{sa12}) and (\ref{p7}), the Bernouilli function is given by
\begin{equation}
\label{p10}
B={1\over\beta}\ln Z,
\end{equation}
and it plays the role of a free energy in the statistical theory (if we interpret $Z$ as a partition function).

We also note that the
most probable PV field $\langle \sigma\rangle({\bf r})$ of the Gibbs
distribution (\ref{p5}) is given by \citep{nic}:
\begin{equation}
\label{p11}
\langle \sigma\rangle=\lbrack (\ln\chi)'\rbrack^{-1}(\beta\psi+\alpha),
\end{equation}
provided that $(\ln\chi)''(\langle\sigma\rangle)<0$.  This is also a
stationary solution of the SW system which usually differs
from the average value $\overline{q}({\bf r})$ of the Gibbs
distribution (\ref{p5}). They coincide only when
\begin{equation}
\label{p12}
-(\ln\hat\chi)'(\Phi)=\lbrack (\ln\chi)'\rbrack^{-1}(\Phi),
\end{equation}
which is the case  when $\chi(\sigma)$ is gaussian.

\subsection{Generalized entropy}
\label{ge}

We first show that a PV field which extremizes a functional of the form
\begin{equation}
H[{q}]=-\int C({q})h d{\bf r},
\label{ge1}
\end{equation}
where $C$ is a convex function, at fixed energy $E$, mass $M$ and circulation
$\Gamma$, is a steady state of the SW equations. We write the variational
principle as
\begin{equation}
\delta H-\beta\delta E-\alpha \delta\Gamma-\gamma\delta M=0.
\label{ge2}
\end{equation}
Using the results of \citet{sw}, we have
\begin{equation}
\delta E=\int B\delta h d{\bf r}+\int \psi \delta({q}h)d{\bf r}-\int \phi\delta(\nabla\cdot {\bf u})d{\bf r},
\label{ge3}
\end{equation}
\begin{equation}
\delta \Gamma=\int \delta({q}h)d{\bf r},
\label{ge4}
\end{equation}
\begin{equation}
\delta H=-\int C({q})\delta h d{\bf r}-\int C'({q})  \delta ({q}h) d{\bf r}+\int C'({q}) {q} \delta h d{\bf r},
\label{ge5}
\end{equation}
where we have taken $h$, ${q}h$ and $\nabla\cdot {\bf u}$ as independent
variables.
The variations on $\nabla\cdot {\bf u}$ yield $\phi=0$. The variations on ${q}h$ give
\begin{equation}
C'({q})=-\beta\psi-\alpha,
\label{ge6}
\end{equation}
so that ${q}=f(\psi)$. The variations on $h$ give
\begin{equation}
{q}C'({q})-C({q})=\beta B+\gamma,
\label{ge7}
\end{equation}
so that $B=B(\psi)$. Taking the derivative of Eq. (\ref{ge7}) with
respect to $\psi$, we find that
\begin{equation}
{q}C''({q}){d{q}\over d\psi}=\beta {dB\over d\psi}.
\label{ge8}
\end{equation}
According to Eq. (\ref{ge6}), we also have
\begin{equation}
{d{q}\over d\psi}=-{\beta\over C''({q})},
\label{ge9}
\end{equation}
so that $q(\psi)$ is a monotonic function increasing at negative
temperatures and decreasing at positive temperatures. Furthermore,
we find that Eq. (\ref{ge8}) is equivalent to ${q}=-dB/d\psi$.
Therefore, the optimization problem (\ref{ge2}) determines
stationary solutions of the SW system. Since $H$, $E$, $\Gamma$ and
$M$ are conserved by the SW equations, we can argue, as for the 2D
Euler equation \citep{ellis}, that a {\it maximum} of $H$ at fixed
$E$, $\Gamma$ and $M$ (if it exists) will be nonlinearly dynamically
stable with respect to the SW system. In this dynamical context, $H$
is referred to as a Casimir functional and $E-H$ as an
energy-Casimir functional.

Note that the optimization problem (\ref{ge2}) can also be justified
by a selective decay principle (for $-H$) if we interpret the PV
vorticity as the {\it coarse-grained} PV. Indeed, $-H[\overline{q}]$
calculated with the coarse-grained PV decreases (fragile constraint)
while $E[\overline{q}]$, $\Gamma[\overline{q}]$ and
$M[\overline{q}]$ are approximately conserved (robust constraints).
This {\it selective decay principle} can explain physically {\it
how} $-H[\overline{q}]$ can possibly reach a minimum value while
$-H[{q}]$ is exactly conserved by the SW equations. In this
coarse-grained context, $H[\overline{q}]$ is referred to as a
generalized $H$-function \citep{tremaine}.

Finally, we note that the equilibrium state (\ref{p7}) predicted by the
statistical theory extremizes a certain $H$-function at fixed $E$,
$\Gamma$ and $M$. This functional
\begin{equation}
S[\overline{q}]=-\int C(\overline{q})h d{\bf r},
\label{ge10}
\end{equation}
corresponding to the statistical equilibrium state, will be called a
generalized entropy in $\overline{q}$-space \citep{super}. It is completely
determined by the prior PV distribution. It should not be confused
with the mixing entropy (\ref{free1}) which is a functional of $\rho$. Coming
back to Eq. (\ref{ge6}) and recalling that $C$ is convex (so that this relation can be inversed), we find that Eqs. (\ref{ge6}) and
(\ref{p7}) coincide provided that
$C'(\overline{q})=-F^{-1}(\overline{q})$. Therefore, the prior PV
distribution $\chi(\sigma)$ determines $F(x)$ which itself determines
$C(\overline{q})$ according to
\begin{equation}
C(\overline{q})=-\int^{\overline{q}}F^{-1}(x)dx.
\label{ge11}
\end{equation}
In other words, to obtain $C(\overline{q})$, we need to inverse Eq. (\ref{p7}) and integrate the resulting expression.  Some examples are collected in \citet{pre}. Combining the previous relations, we find that the generalized entropy is determined by the prior according to \citep{super}
\begin{equation}
\label{ge12} C(\overline{q})=-\int^{\overline{q}}\lbrack (\ln {\hat \chi})'\rbrack^{-1}(-x)dx.
\end{equation}
 Finally, comparing Eqs. (\ref{ge9}) and (\ref{p9}) we get the relation
\begin{equation}
q_{2}={1\over C''(\overline{q})},
\label{ge13}
\end{equation}
which is exact at statistical equilibrium.

\section{Relaxation equations}

\subsection{Maximum Entropy Production Principle}
\label{mepp}

In the case of freely evolving flows, \citet{sw} have proposed a
thermodynamical parametrization of the SW equations (on the
coarse-grained scale) in the form of relaxation equations that
conserve all the constraints of the SW dynamics (including the
Casimirs)  and increase the mixing entropy. In the case of flows
that are forced at small-scales, the Casimirs are replaced by the
specification of a prior vorticity distribution or, as we have seen,
by a generalized entropy. In this context, we can propose a
thermodynamical parametrization of the SW equations in the form of
relaxation equations that conserve only the robust constraints
(mass, energy and circulation) and increase the generalized entropy
$S[\overline{q}]$ fixed by the prior.

We first decompose the vorticity $\omega$ and velocity $\bf{u}$ into a
mean and fluctuating part, namely $\omega
=\overline{\omega}+\tilde{\omega }$,
${\bf{u}}=\overline{\bf{u}}+\tilde{\bf{u}}$, keeping $h$
smooth. Taking the local average of the shallow water equations
(\ref{sa1})(\ref{sa3}), we get
\begin{equation}
{\partial h\over\partial t}+\nabla \cdot (h\overline{\bf u})=0,
\label{m1}
\end{equation}
\begin{equation}
{\partial \overline{\bf u}\over\partial t}+(\overline{\mb\omega}+2{\bf\Omega})\times \overline{\bf u}=-\nabla B-{\bf e}_{z}\times {\bf J}_{\omega},
\label{m2}
\end{equation}
where the current $\bf{J}_{\omega }=\overline{\tilde{\omega
}\tilde{\bf{u}}}$ represents the correlations of the fine-grained
fluctuations.  We deduce an equation for the evolution of the
potential vorticity (\ref{sa5}), taking the curl of Eq. (\ref{m2})
and using Eq. (\ref{m1}):
\begin{equation}
{\partial\over \partial t}(h\overline{q})+\nabla\cdot (h\overline{q}\ \overline{\bf u})=-\nabla\cdot {\bf J}_{\omega}.
\label{m3}
\end{equation}
This equation can be viewed as a local conservation law for the
circulation $\Gamma =\int \overline{q}hd{\bf{r}}$. It shows also
that $ {\bf J}_{\omega}$ represents the current of coarse-grained
vorticity due to mixing. We shall determine the unknown current
${\bf{J}}_{\omega }$ by the thermodynamic principle of Maximum
Entropy Production (MEP), using the generalized entropy
(\ref{ge10}). The Maximum Entropy Production (MEP) principle
consists in choosing the current ${\bf{J}}_{\omega}$ which maximizes
the rate of entropy production $\dot{S}$ respecting the constraints
$\dot{E}=0$, and $J_{\omega}^{2}\le C({\bf r},t)$. The last
constraint expresses a bound (unknown) on the value of the diffusion
current. Convexity arguments justify that this bound is always
reached so that the inequality can be replaced by an equality. We
write the variational problem as
\begin{equation}
\delta \dot S-\beta(t)\delta \dot E-\int D^{-1}\delta\biggl
({J_{\omega}^{2}\over 2}\biggr )d{\bf r}=0, \label{m4}
\end{equation}
where $\beta(t)$ is a Lagrange multiplier accounting for the
conservation of energy and $D^{-1}$ is a Lagrange multiplier
associated with the constraint $J_{\omega}^{2}= C({\bf r},t)$. The
conservations of mass and circulation are automatically satisfied by
the form of the relaxation equations (\ref{m1}) and (\ref{m2}). Noting
that
\begin{equation}
\dot E=\int {\bf J}_{\omega}\cdot \overline{{\bf u}}_{\perp} h d{\bf r},
\label{m5}
\end{equation}
\begin{equation}
\dot S=-\int C''(\overline{q}){\bf J}_{\omega}\cdot \nabla \overline{q} d{\bf r},
\label{m6}
\end{equation}
and performing the variations on ${\bf J}_{\omega}$ in Eq.
(\ref{m4}), we obtain an optimal current
\begin{equation}
{\bf J}_{\omega}=-D\biggl \lbrack \nabla \overline{q}+{\beta(t)\over C''(\overline{q})}h \overline{{\bf u}}_{\perp}\biggr\rbrack.
\label{m7}
\end{equation}
Thus, in the presence of a prior PV distribution, we obtain a
parametrization of the SW equations of the form
\begin{equation}
\label{m8}
{\partial h\over \partial t}+\nabla \cdot (h{\mathbf{u}})=0,
\end{equation}
\begin{equation}
\label{m9}
{\partial \mathbf{u}\over \partial t}+\overline{q}h\mathbf{e}_{{z}}\times \mathbf{u}=-\nabla \biggl ({\mathbf{u}^{2}\over 2}+gh\biggr )+D\left\lbrack{\mathbf{e}}_{z}\times \nabla \overline{q}-{\beta (t)\over C''(\overline{q})}h\mathbf{u}\right\rbrack,
\end{equation}
\begin{equation}
\label{m10}
\overline{q}={{(\nabla \times \mathbf{u})\cdot\mathbf{e}_{{z}}+2\Omega }\over h}\; \; ,\; \; \beta (t)=-{\int Dh{\textbf {u}}_{\perp }\cdot \nabla \overline{q}d{\mathbf{r}}\over \int D{{\textbf {u}}^{2}h^{2}\over C''(\overline{q})}d{\mathbf{r}}},
\end{equation}
\begin{equation}
\label{m11} {\textbf {n}}\cdot \nabla \overline{q}=-{\beta (t)\over
C''(\overline{q})}h\ {\textbf {n}}\cdot {\textbf {u}}_{\perp }\qquad
({\textrm{on }\, \, \textrm{each }\, \, \textrm{boundary}}),
\end{equation}
\begin{equation}
\label{m12}
{\textbf {n}}\cdot {\textbf {u}}=0\qquad ({\textrm{on }\, \, \textrm{each }\, \, \textrm{boundary}}),
\end{equation}
where ${\bf n}$ is a unit vector normal to the boundary and we have
omitted the overbar on ${\bf u}$. The relaxation equation for the
coarse-grained vorticity is given by
\begin{equation}
{\partial\over \partial t}(h\overline{q})+\nabla\cdot (h\overline{q}{\bf u})=\nabla\cdot \left\lbrace D\biggl \lbrack \nabla \overline{q}+{\beta(t)\over C''(\overline{q})}h \overline{{\bf u}}_{\perp}\biggr\rbrack\right\rbrace.
\label{m13}
\end{equation}
These equations can also be directly obtained from the
parametrization of \citet{sw} by replacing $q_{2}$ in their
parametrization by $1/C''(q)$. Therefore, the identity (\ref{ge13})
can be viewed as a {\it closure relationship} in the present
context. This relation is valid at equilibrium but the present
approach suggests  that it is also valid out-of-equilibrium when
there is a prior distribution of PV. In fact, we can obtain this
relation by assuming that, out-of-equilibrium, the PV distribution
$\rho({\bf r},\sigma,t)$ maximizes the relative entropy (\ref{p4})
at fixed PV $\overline{q}({\bf r},t)$ and normalization (Appendix C
of \citet{physicad}).

The entropy production (\ref{m6}) can be written
\begin{equation}
\dot S=-\int C''(\overline{q}){\bf J}_{\omega}\cdot \left\lbrack \nabla \overline{q}+{\beta(t)\over C''(\overline{q})}h{\bf u}_{\perp}\right\rbrack d{\bf r}+\beta(t)\int {\bf J}_{\omega}\cdot {\bf u}_{\perp} h d{\bf r}.
\label{m14}
\end{equation}
Using the conservation of energy $\dot E=0$ with Eq. (\ref{m5}), the second integral is seen to vanish. Substituting Eq. (\ref{m7}) in the first integral, we finally obtain
\begin{equation}
\dot S=\int C''(\overline{q}){{\bf J}_{\omega}^{2}\over D} d{\bf r}\ge 0,
\label{m15}
\end{equation}
which is positive provided that $D\ge 0$. A stationary solution of
the relaxation equations (\ref{m8})-(\ref{m12}) satisfies $\dot S=0$
yielding ${\bf J}_{\omega}={\bf 0}$, i.e.
\begin{equation}
\nabla\overline{q}+{\beta\over C''(\overline{q})}\nabla\psi={\bf 0}.
\label{m16}
\end{equation}
After integration, we obtain
\begin{equation}
C'({q})=-\beta\psi-\alpha.
\label{m17}
\end{equation}
Therefore, the generalized entropy increases until the statistical
equilibrium state (\ref{p7})-(\ref{ge6}), fixed by the prior
$\chi(\sigma)$, is reached with $\beta=\lim_{t\rightarrow
+\infty}\beta(t)$. Alternatively, these equations can be used as a
numerical algorithm to compute arbitrary stationary solutions of the
SW system specified by the convex function $C$ (see Sec. \ref{ge}).

\subsection{The incompressible limit}

The case of ordinary 2D incompressible turbulence is recovered in the
limit $h\rightarrow 1$, $q\rightarrow \omega$ and ${\textbf
{u}}=-{\textbf {e}}_{z}\times \nabla \psi$. The relaxation equation for the coarse-grained vorticity is given by
\begin{equation}
{\partial\overline{\omega}\over \partial t}+{\bf u}\cdot \nabla\overline{\omega}=\nabla\cdot \left\lbrace D\biggl \lbrack \nabla \overline{\omega}+{\beta(t)\over C''(\overline{\omega})}\nabla\psi\biggr\rbrack\right\rbrace,
\label{inc1}
\end{equation}
with
\begin{equation}
\label{inc2}
\beta (t)=-{\int D\nabla \overline{\omega}\cdot\nabla\psi d{\mathbf{r}}\over \int D{(\nabla\psi)^{2}\over C''(\overline{\omega})}d{\mathbf{r}}}.
\end{equation}
This returns the equations introduced by \citet{pre} in 2D turbulence in the case where the system is described by a prior vorticity distribution. They can be viewed as nonlinear mean-field Fokker-Planck equations. They are the forced-case  counterpart of the  relaxation equations introduced by \citet{rs2} for freely evolving flows that conserve all the Casimir constraints.

The relaxation equation (\ref{m9}) for the velocity field can be written
\begin{equation}
\label{inc3}
{\partial {{\bf u}}\over \partial t}+({{\bf u}}\cdot \nabla ){{\bf u}}=-{1\over \rho }\nabla p+D\left \lbrack\Delta {{\bf u}}-{\beta (t)\over C''(\overline{\omega})}{{\bf u}}\right \rbrack,
\end{equation}
where we have used the well-known identity of vector analysis $\Delta
{\mathbf{u}}=\nabla (\nabla\cdot {\mathbf{u}})-\nabla \times (\nabla
\times {\mathbf{u}})$ which reduces for a two-dimensional
incompressible flow to $\Delta {\mathbf{u}}={{\mathbf{e}}}_{z}\times
\nabla \overline{\omega }$.  Eq. (\ref{inc3}) is valid even if $D$
is space dependant unlike with a usual viscosity term. In previous
publications this equation was given only in its vorticity form
(\ref{inc1}) and the equivalence with Eq. (\ref{inc3}) is not obvious
at first sights when $D$ is space dependent. At equilibrium, we have
from Eq. (\ref{inc3}) the identity
\begin{equation}
\label{inc4}
\Delta {\mathbf{u}}={\beta\over C''(\overline{\omega})} {\mathbf{u}},
\end{equation}
which can be deduced directly from the stationary state (\ref{ge6}). Indeed,
for a stationary solution $\overline{\omega }=f(\psi )$, the previous
identity $\Delta {\textbf {u}}={\textbf {e}}_{z}\times
\nabla \overline{\omega }$ becomes $\Delta {\mathbf{u}}=-f'(\psi ){\mathbf{u}}$ which is equivalent to Eq. (\ref{inc4}) for a steady state thanks to
Eq. (\ref{ge9}). Finally, the previous equations  can be extended to the quasi-geostrophic (QG) limit if we replace $\omega$ by the PV $q$ related to the stream-function by
 \begin{equation}
\label{inc5}
q=-\Delta\psi+{\psi\over L_{R}^{2}},
\end{equation}
where $L_{R}$ is the Rossby radius.

\subsection{Explicit examples}

In the present formalism, one has to specify (i) a prior PV
distribution $\chi(\sigma)$ which encodes small-scale forcing and
non-ideal effects (ii) the robust constraints $E$, $\Gamma$ and $M$
which can be determined at any time from the coarse-grained flow.
From the prior PV distribution, we can determine a generalized
entropy $C(\overline{q})$ by using the procedure exposed in Sec.
\ref{ge} (see in particular Eq. (\ref{ge12})). Then, we can 
use this entropy in the parametrization
(\ref{m8})-(\ref{m12}) to obtain the dynamical evolution of the
coarse-grained flow towards statistical equilibrium. The difficulty is
now to find the good prior. This depends from case to case as it is
related to the properties of forcing, but the idea is that several
forms of prior (or corresponding entropies) give similar results so
that they can be regrouped in {\it classes of equivalence}
\citep{pre}. Thus, for a given situation, one has to find the relevant
class of equivalence. In general, one has to proceed by tryings and
errors. We specify a prior, compute the corresponding flow and see
whether it agrees with the information that we have on the system. If
we find a ``good prior'' for a system with some given initial
conditions, then we can expect that it will remain valid for this
system when we change the initial conditions (or, equivalently, the
value of the robust constraints $E$, $\Gamma$ and $M$). Note that
specifying the prior is not the end of the story but only the starting
point. Indeed, there can be different types of solutions for a given
prior PV distribution depending on the values of the control
parameters.  Thus, we have to study the bifurcation diagram as a
function of these parameters $E$, $\Gamma$ and $M$ for a given prior
$\chi(\sigma)$ or generalized entropy $C(\overline{q})$. This is a
rich and non trivial problem.

Let us give some examples to illustrate our approach. In the case of Jovian flows, \citet{ellis} have proposed to adopt a prior $\chi(\sigma)$ corresponding to a de-centered Gamma distribution. This leads to a generalized entropy of the form \citep{super}
\begin{equation}
\label{id1}
C(\overline{q})={1\over\epsilon}\left \lbrack \overline{q}-{1\over\epsilon}\ln (1+\epsilon\overline{q})\right\rbrack,
\end{equation}
where $2\epsilon$ is the skewness of the PV distribution. We have $C'(\overline{q})=\overline{q}/(1+\epsilon\overline{q})$ and $C''(\overline{q})=1/(1+\epsilon\overline{q})^{2}$ so that the statistical equilibrium state is specified by
\begin{equation}
\label{id2}
\overline{q}=-{\beta\psi+\alpha\over 1+\epsilon (\beta\psi+\alpha)}.
\end{equation}
In the limit $\epsilon\rightarrow 0$, the generalized entropy
(\ref{id1}) becomes minus the enstrophy $S[\overline{q}]=-{1\over
2}\int \overline{q}^{2}d{\bf r}$ and the $\overline{q}=f(\psi)$
relationship is linear. The parametrization that we propose in that
case for the vorticity equation is
\begin{equation}
{\partial\over \partial t}(h\overline{q})+\nabla\cdot (h\overline{q}{\bf u})=\nabla\cdot \left\lbrace D\biggl \lbrack \nabla \overline{q}+\beta(t)(1+\epsilon\overline{q})^{2}  h \overline{{\bf u}}_{\perp}\biggr\rbrack\right\rbrace.
\label{id3}
\end{equation}

On the other hand, in order to describe jovian flows, \citet{nore}
have considered a case of the statistical theory where the PV
distribution is restricted to two-levels. In their approach, the flow
is assumed to be freely evolving and the dynamics corresponds to the
inviscid mixing of patches with PV $\sigma_{0}$ and $\sigma_{1}$. As
discussed in
\citet{physicad}, we can also consider the case of a flow that is
permanently forced at small-scales (due to convection) so that a
prior PV distribution is established with two intense peaks at
$\sigma_{0}$ and $\sigma_{1}$. In the two-levels case, theses two
interpretations lead to the same results but the second one can
probably be extended more easily to the more realistic situation
where the peaks have a finite width $\Delta\sigma$. If we adopt a
prior of the form
$\chi(\sigma)=\delta(\sigma-\sigma_{1})+\chi\delta(\sigma-\sigma_{0})$,
we find that the generalized entropy is
\begin{equation}
\label{id4}
C(\overline{q})={1\over \sigma_{1}-\sigma_{0}}\left\lbrack (\overline{q}-\sigma_{0})\ln (\overline{q}-\sigma_{0}) +(\sigma_{1}-\overline{q})\ln(\sigma_{1}-\overline{q})\right\rbrack.
\end{equation}
We have
$C'(\overline{q})={1\over\sigma_{1}-\sigma_{0}}\ln({\overline{q}-\sigma_{0}\over
\sigma_{1}-\overline{q}})$ and $C''(\overline{q})=1/\lbrack
({\overline{q}-\sigma_{0})(\sigma_{1}-\overline{q}})\rbrack$ so that
the statistical equilibrium state is
\begin{equation}
\label{id5} \overline{q}=\sigma_{0}+{\sigma_{1}-\sigma_{0}\over
1+\chi e^{(\sigma_{1}-\sigma_{0})(\beta\psi+\alpha)}},
\end{equation}
similar to the Fermi-Dirac distribution. The parametrization that we
propose in that case for the vorticity equation is
\begin{equation}
{\partial\over \partial t}(h\overline{q})+\nabla\cdot (h\overline{q}{\bf u})=\nabla\cdot \left\lbrace D\biggl \lbrack \nabla \overline{q}+\beta(t) ({\overline{q}-\sigma_{0})(\sigma_{1}-\overline{q}})  h \overline{{\bf u}}_{\perp}\biggr\rbrack\right\rbrace,
\label{id6}
\end{equation}
and it coincides with the parametrization of \cite{sw} in the
two-levels case of the statistical theory. Therefore, the approach
based on priors is not in radical opposition with the usual
statistical theory but it allows for convenient extensions in
more general cases.

Explicit determinations of the statistical equilibrium state specified
by Eqs. (\ref{id2}) and (\ref{id5}) have been obtained in the QG limit
of the statistical theory. In the two approaches, a vortex solution
has been found at the latitude of Jupiter's great red spot where the
underlying topography is extremum (or the shear is equal to zero). In
the case of the Fermi-Dirac distribution (\ref{id5}), \citet{nore}
consider the limit of small Rossby radius and find that the vortex has
an annular jet structure which is consistent with the morphology of
Jupiter's great red spot. This study has been further developed in
\citet{bs} and it has been extended by \citet{bcs} to the case of the
SW system. In these studies, the structure of Jupiter's great red spot
can be seen as the co-existence of two thermodynamical phases in
contact separated by a sort of ``domain wall'' \citep{physicad} in the
langage of phase ordering kinetics (like in the Cahn-Hilliard
theory). This annular jet structure is not obtained in the approach of
\cite{ellis}. This implies that the prior PV distribution relevant to
the case of JGRS is probably closer to two intense peaks rather than
to a decentered Gamma distribution. These two distributions belong to
different classes of equivalence since they lead to structurally
different solutions. Probably, the prior distribution could be
improved to give a better description of jovian vortices, but a
distribution with two intense peaks already gives a fair description.

\section{Conclusion}

In this paper, we have extended the ``ordinary'' statistical theory of
the SW system \citep{sw} to account for the existence of a prior
vorticity distribution. This approach can be justified when the system
is forced at small-scales so that a permanent PV distribution is
imposed {\it canonically}. In that case, the forcing acts like a sort
of reservoir. The attractive nature of this theory is its {\it
practical interest}: in the standard theory, one works with the
Boltzmann entropy in $\rho$-space and deals with a very large
(possibly infinite) number of Casimir constraints which are often not
known or not rigorously conserved. In the other approach, one
conserves only the robust constraints ($E$, $\Gamma$, $M$) and work
with a generalized entropic functional in $\overline{q}$-space fixed
by the prior (that has to be found by tryings and errors). Therefore,
in the first approach, we have to solve $N$ coupled relaxation
equations (one for each level) while in the second approach, we just
have to solve one relaxation equation.  Whether this approach is
really physically relevant remains to be established. In any case, the
relaxation equations (\ref{m8})-(\ref{m12}) can be used as numerical
algorithms to construct a large class of stationary solutions of the
SW equations, which is certainly an interest of our formalism.

\end{document}